\documentclass[twocolumn,prb,showpacs,superscriptaddress]{revtex4}
\usepackage{graphics}
\usepackage{graphicx}
\usepackage{amsmath}
\usepackage{amssymb}
\usepackage{subfigure}
\usepackage{longtable}
\usepackage{multirow}
\usepackage{bm}
\usepackage{hyperref}
\usepackage{multirow}

\newcommand{\ve}[1]{\bm{\mathrm{#1}}} 
\newcommand{\rr}{\mathbf{r}}

\newcommand{\np}{n^{\prime}}
\newcommand{\w}{\omega}
\newcommand{\kk}{\mathbf{k}}
\newcommand{\kq}{\mathbf{k} + \mathbf{q}}

\begin{document}
\title{Optical properties of bulk semiconductors and graphene/boron-nitride: The Bethe-Salpeter equation with derivative discontinuity-corrected DFT energies}  

\author{Jun Yan}
\email{junyan@stanford.edu}
\affiliation{Center for Atomic-scale Materials Design, Department of
Physics \\ Technical University of Denmark, DK - 2800 Kgs. Lyngby, Denmark}
\affiliation{SUNCAT Center for Interface Science and Catalysis,  
SLAC National Accelerator Laboratory \\ 2575 Sand Hill Road, Menlo Park, CA 94025, USA}
\author{Karsten W. Jacobsen}
\affiliation{Center for Atomic-scale Materials Design, Department of
Physics \\ Technical University of Denmark, DK - 2800 Kgs. Lyngby,
Denmark}
\author{Kristian S. Thygesen}
\affiliation{Center for Atomic-scale Materials Design, Department of
Physics \\ Technical University of Denmark, DK - 2800 Kgs. Lyngby, Denmark}
\affiliation{Center for Nanostructured Graphene (CNG), Department of Micro- and Nanotechnology \\ DTU Nanotech,
Technical University of Denmark, DK - 2800 Kgs. Lyngby,
Denmark}

\date{\today}

\begin{abstract}
  We present an efficient implementation of the Bethe-Salpeter
   equation (BSE) for optical properties of materials in the projector
  augmented wave method GPAW. Single-particle energies and wave
  functions are obtained from the GLLBSC functional which explicitly
  includes the derivative discontinuity, is computationally inexpensive, and yields excellent fundamental gaps. Electron-hole
  interactions are included through the BSE using the statically
  screened interaction evaluated in the random phase approximation. For a representative set of semiconductors and insulators we find excellent agreement with experiments for the dielectric
  functions, onset of absorption, and lowest excitonic features. For
  the two-dimensional systems of graphene and hexagonal boron-nitride
  (h-BN) we find good agreement with previous many-body calculations. For
  the graphene/h-BN interface, we find that the fundamental and optical
  gaps of the h-BN layer are reduced by 2.0 eV and 0.7 eV, respectively,
  compared to freestanding h-BN. This reduction is due to image charge
  screening which shows up in the GLLBSC calculation as a reduction (vanishing)
  of the derivative discontinuity.

\end{abstract}

\maketitle

\section{Introduction}
Optical spectroscopies such as photo absorption, luminescence, and
reflectance measurements are widely used for materials
characterization. In this context, first-principles calculations play an increasingly important role for the interpretation and guidance of experimental investigations. 
However, theoretical spectroscopic methods are not only useful for characterization purposes. Indeed, with the recent focus on solar energy conversion,
plasmonics, and optoelectronics -- all applications which involve the interaction of light with matter -- first-principles methods for calculating the optical properties of complex materials are becoming the essential tool allowing for reliable computational design of new materials within these areas.

The two most commonly used {\it ab-initio} methods for optical properties are time-dependent density functional theory (TDDFT)\cite{Gross_L84} and
many-body perturbation theory (MBPT)\cite{Rubio_RMP02}. For smaller molecules
and clusters\cite{Casida_98}, TDDFT with the adiabatic local density approximation
(ALDA) provides a reasonably good compromise between accuracy and
computational cost.  However, the ALDA fails to describe several
important effects including the formation of excitons in extended systems\cite{TDDFT_review}, 
charge-transfer excitations in donor-acceptor molecular complexes\cite{Gordon_JCP03,Juanma_L11}, as well as the
screening of optical transitions by nearby metal surfaces\cite{Juanma_L11}. Apart from these qualitative failures, the ALDA is also found to underestimate the optical transition energies and overestimate static dielectric constants of bulk insulators and semiconductors. This problem is, at least to some extent, related to the well known tendency
of the LDA and related semi local exchange-correlation (xc) functionals, to underestimate the
fundamental energy gaps in such systems. 

All of the above mentioned problems of the TDDFT-ALDA approach are overcomed by the
MBPT.  In the standard scheme, the quasiparticle band structures are
obtained using the GW approximation\cite{Aryasetiawan_GWreview} while optical excitation energies
are obtained by solving a Bethe-Salpeter equation (BSE)\cite{BS_51} with a
statically screened electron-hole interaction. The GW-BSE approach\cite{Reining_L98_Si,Louie_B00} has
been succesfully applied to a number of different systems ranging from
bulk semiconductors\cite{Reining_L98_Si}, insulators and their surfaces\cite{Louie_L99_Sisurface}, two-dimensional systems such
as graphene\cite{Louie_L09_graphene} and boron nitride layers\cite{Alouani_L06_bulkBN}, metal-molecule
interfaces\cite{Juanma_L11}, isolated molecules\cite{Reining_L95_Na4, Louie_L98_SiH,Chelikowsky_L06_CdSe} and liquid water\cite{Reining_L06_water}.  Nevertheless, applications of the
approach to larger systems are limited by the extremely demanding
computational requirements of both the GW and BSE calculations. 

Several schemes have been proposed to reduce the computational cost of
GW-BSE calculations. These include circumventing the GW step by
applying simpler band structures e.g. derived from the COHSEX
approximation\cite{Hedin_1965} or simply scissors operator-corrected LDA band structures\cite{Levine_L89}, or the
use of model dielectric functions to describe the screening\cite{Louie_B86_GW}. Another
route of research is directed towards the development of more accurate
TDDFT xc-kernels without sacrificing the computational simplicity
associated with this approach\cite{Marini_L03_fxc, Ullrich_B09_fxc, Gross_L11}.

Recently, Kuisma \emph{et al.} have introduced the GLLBSC
xc-potential\cite{gllbsc} which is based on an earlier functional developed by Gritsenko \emph{et al.}\cite{gllb}. This potential explicitly includes the derivative
discontinuity of the xc-potential at integer particle numbers which is
important to obtain physically meaningful band gaps from DFT. 
The derivative discontinuity, $\Delta_{\mathrm{xc}}$,
is calculated directly from the Kohn-Sham eigenvalues and eigenstates. The fundamental band
gap is then obtained as the sum of the Kohn-Sham single-particle gap and
the derivative discontinuity.  The GLLBSC method has been shown to
produce fundamental band gaps as well as band dispersions for a range
of semiconductors in very good agreement with experiments and more
sophisticated theoretical approaches while the computational cost is
comparable to that of LDA\cite{gllbsc,Ivano,Jun_AgH}.

In this paper we combine the TDDFT and BSE methods for treating the
electron-hole interaction with the GLLBSC method for the wave function
and band structures. Considering both bulk and low dimensional systems
we find that the accuracy of the GLLBSC-BSE approach is comparable to
the GW-BSE approach. All the methods are implemented in the
\textsc{gpaw} code\cite{GPAW, GPAW_10, Jun_response}, an electronic structure package
based on the projector augmented wave methodology\cite{Blochl_B94,
  Blochl_03}. For the bulk systems Si, C, InP, MgO, GaAs and LiF, we
find that the fundamental gaps and static dielectric constants
calculated with GLLBSC compare well with experimental data.
Importantly, the static dielectric constant should be evaluated
without the derivative discontinuity when using an xc-kernel that does
not account for e-h interaction such as the ALDA or the random phase
approximation (RPA). The experimental optical absorption spectra of
all compounds are also very well reproduced by the GLLBSC-BSE approach
including the absorption onset and excitonic peaks. Finally, the
method is used to compute the band structure and optical absorption
spectra of graphene, hexagonal boron-nitride (h-BN), and a
graphene/h-BN interface. For the isolated sheets we find good
agreement with previous GW-BSE calculations. For the interface we find
that both the quasiparticle- and optical gap of the h-BN sheet are
reduced by 2.0 and 0.7 eV, respectively. The physical origin of this effect is due to image charge screening
by the graphene layer. In the GLLBSC, the reduction shows up as a vanishing of the derivative discontinuity.

The rest of the paper is organized as follows. Section II introduces
the theoretical framework for calculating optical properties of solids
with \textsc{gpaw} using the TDDFT and BSE approaches, followed by a
brief review of the GLLBSC method.  Details of the implementation are
presented in Sec. III. Section IV presents benchmark results for the
band gaps, dielectric constants and optical absorption spectra of a
number of bulk semiconductors and insulators. In Sec. V we present the
band structures and optical spectra of graphene, h-BN, and
graphene/h-BN interface. Finally, a summary is given in Sec. V.

\section{Method}
\subsection{Macroscopic dielectric function}
Most of the optical properties of a solid can be obtained from the
macroscopic dielectric function,

\begin{equation}
\label{Eq:epsilon_M}
  \epsilon(\omega) \equiv \left. \frac{1}{\epsilon^{-1}_{\ve G \ve G^{\prime}}(\ve q\rightarrow 0, \w)} \right |_{\ve G=0, \ve G^{\prime}=0}. 
\end{equation}
Here, $\epsilon_{\ve G \ve G^{\prime}} (\ve q, \w)$ is the (microscopic) dielectric
matrix in reciprocal $\ve G$ space. The off-diagonal elements of the
$\epsilon$ matrix account for local field effects arising due to the
periodic crystal potential. The macroscopic average is achieved
through the inversion of the $\epsilon$ matrix. 

In this work we consider only the longitudinal component of the
dielectric function. For applications to optical properties this is in fact not a restriction because in the relevant long wave length limit the electrons do not feel the
difference between longitudinal and transversely polarized fields, and
consequently the two types of response functions coincide.
Still, for anisotropic systems $\epsilon(\omega)$ depends on the direction in
which the limit $\ve q\to 0$ is taken. However, to keep the notation simple we
shall omit reference to this direction in what follows.

\subsection{Linear response function from TDDFT}
The microscopic dielectric matrix is related to the linear density response function, $\chi$, via
\begin{equation}
\label{Eq:epsilon_-1}
 \epsilon^{-1}_{\mathbf G \mathbf G^{\prime}}(\mathbf q, \omega)
  = \delta_{\mathbf G \mathbf G^{\prime}} + \frac{4\pi}{|\mathbf q + \mathbf G| |\mathbf q + \mathbf G^{\prime}|} 
  \chi_{\mathbf G \mathbf G^{\prime}}(\mathbf q, \omega). 
\end{equation}
Within TDDFT the response function is related to the response function of the non-interacting Kohn-Sham electrons, $\chi^0$ and the exchange-correlation interaction kernel $K_{xc}$ via
a Dyson-like equation, 
\begin{eqnarray}
\label{Eq: Dyson_GG}
 & \chi_{\mathbf G \mathbf G^{\prime}}(\mathbf q, \omega)  
  = \chi^0_{\mathbf G \mathbf G^{\prime}}(\mathbf q, \omega) \nonumber \\
 & + \sum_{\mathbf G_1 \mathbf G_2} \chi^0_{\mathbf G \mathbf G_1}(\mathbf q,
\omega) K_{\mathbf G_1 \mathbf G_2}(\mathbf q,\omega)
  \chi_{\mathbf G_2 \mathbf G^{\prime}}(\mathbf q, \omega). 
\end{eqnarray}
The KS response function is given by\cite{Adler, Wiser},  
\begin{align}\label{Eq:chi0_GG}
& \chi^0_{\mathbf{G} \mathbf{G}^{\prime}}(\mathbf{q}, \omega) = \frac{2}{\Omega} 
 \sum_{\ve k, n n^{\prime}} (f_{n\mathbf{k}}-f_{n^{\prime} \mathbf{k}+\ve q
}) \nonumber\\
&  \times  
 \frac{n_{n\ve k,\np \kq}(\ve G) n^{\ast}_{n\ve k,\np \kq}(\ve G^{\prime})}{\omega +
\epsilon_{n\mathbf{k}} - \epsilon_{n^{\prime} \mathbf{k}+\ve q } + i\eta}
\end{align} 
where $\varepsilon_{n \mathbf{k}}$ is a KS eigenvalue, and $f_{n\mathbf{k}}$ is the occupation factor. The quantity 
\begin{equation}
\label{Eq:pair_density}
 n_{n\ve k,\np \kq}(\ve G) \equiv 
\langle \psi_{n\kk} | e^{-i (\ve q + \ve G) \cdot \rr} | \psi_{n^{\prime} \ve k +\ve q}
\rangle 
\end{equation}
is referred to as the charge density matrix\cite{Jun_response}. In the long wavelength limit, i.e. for $\ve q \rightarrow 0$, and for $n\neq n'$, application of the $k\cdot p$ perturbation theory\cite{Grosso} yields the 
important identity 
\begin{equation}
\label{Eq:density_matrix}
\mathrm{lim}_{\ve q \rightarrow 0} n_{n\ve k,\np \ve k+\ve q}(0) = \frac{-i \ve q \cdot \langle \psi_{n\kk} | \nabla | \psi_{n^{\prime} \ve k}
\rangle}{\epsilon_{n^{\prime} \ve k}- \epsilon_{n \ve k} }. 
\end{equation}
Alternatively, this form follows directly if we consider the density induced by a longitudinal vector potential rather than a scalar potential.
A detailed description of the evaluation of the charge density matrix
and the ALDA xc-kernel within the PAW formalism can be found in Ref.
\onlinecite{Jun_response}.

\subsection{The Bethe-Salpeter Equation}\label{sec.bse}
Several of the shortcomings of the ALDA in describing optical spectra
are overcomed by explicitly accounting for electron self-energy effects
and electron-hole interactions using many-body perturbation theory. In
the standard GW-BSE approach, the single-particle energies are
evaluated using a self-energy in the GW approximation while the
optical excitation energies are obtained by diagonalizing an effective
two-particle Hamiltonian. In the present work we avoid calculating the
GW self-energy by using single-particle energies obtained from the
efficient GLLBSC functional.

Following the standard approach, the excitation energies corresponding to
an external potential with momentum $\ve q$ 
 can be found by solving an eigenvalue problem of the form
\begin{equation}\label{eq.bse_eig}
\sum_{S'}\mathcal H(\ve q)_{SS'} A_{S'}^\lambda(\ve q) = E^\lambda(\ve q) A^\lambda_{S}(\ve q)
\end{equation}
where $\mathcal H_{SS'}(\ve q)$ is the Bethe-Salpeter effective two-particle Hamiltonian evaluated in a basis of electron-hole states, $\psi_S(\ve r_h,\ve r_e)=\psi_{n \ve k}(\ve r_h)^*\psi_{m \ve k+\ve q}(\ve r_e)$. 
The BSE Hamiltonian reads
\begin{equation}
\mathcal H_{S S^{\prime}}(\ve q) =(\varepsilon_{m \ve k+\ve q}^{QP}-\varepsilon_{n \ve k}^{QP}) \delta_{S S^{\prime}}  - (f_{m \ve k+\ve q}-f_{n \ve k})
K_{S S^{\prime}}(\ve q)
\end{equation}
The kernel consists of an e-h exchange interaction ($V$) and a direct screened e-h attraction ($W$), 
\begin{align}\label{eq.kernel}
K_{S S^{\prime}}(\ve q)=&V_{SS'}(\ve q)- \frac{1}{2}W_{SS'}(\ve q).
\end{align}
The factor 2 accounts for spin. In appendix \ref{app.bse} we give a derivation of the BSE eigenvalue equation and its relation to the dielectric function.

The effective two particle Hamiltonian is most conveniently evaluated in a plane wave basis. In this representation the e-h exchange term reads 
\begin{equation} \label{Eq:V_eh}
V_{SS^{\prime}} (\ve q)
= \frac{4\pi}{\Omega}\sum_{\ve G} 
\frac{n^{\ast}_{n\ve k,m\ve k+\ve q}(\ve G)n_{n'\ve k',m'\ve k'+\ve q}(\ve G)}{|\ve q + \ve G|^2}, 
\end{equation}
If we exclude the $\ve G=0$ component in the sum we obtain the short
range exchange kernel $\bar V$. The difference between $V$ and
$\bar V$ becomes important when the response function is written
in terms of the eigenstates and energies of the BSE Hamiltionian, see
below. To obtain the optical limit $V_{SS^{\prime}} (\ve q \to 0)$ we
use the expression Eq. (\ref{Eq:density_matrix}) to cancel the $1/q^2$
Coulomb divergence appearing in the $\ve G=0$ term. In the evaluation of the remaining terms we use a small finite value for $\mathbf q$ (a value of 0.0001 \AA$^{-1}$ has been used in this work).

The plane wave expression for the e-h direct Coulomb term reads
\begin{align}
\label{Eq:W_SS}
W_{SS^{\prime}} (\ve q)&= \frac{4\pi}{\Omega}\sum_{\ve G \ve G^{\prime}} n^{\ast}_{n \ve k, n' \ve k'}(\ve G)W_{\ve G \ve G'}(\ve k'-\ve k)\nonumber\\
&\times n_{m \ve k+\ve q, m' \ve k'+\ve q}(\ve G^{\prime}),  
\end{align}
where   
\begin{equation}
 W_{\ve G \ve G^{\prime}} (\ve k'-\ve k) = \frac{\epsilon^{-1}_{\ve G \ve G^{\prime}}(\ve k'-\ve k, \w=0)}{|\ve k'-\ve k + \ve G||\ve k'- \ve k + \ve G^{\prime}|} 
\end{equation}
Here we encounter a divergence of $W_{\ve G \ve G'}$ when either $\ve
G$ or $\ve G'$ is zero and $\ve k=\ve k'$. Such a divergence due to the singularity of the Coulomb kernel at $q=0$ is also present in calculating exact exchange\cite{Alavi_B08} and GW self energies\cite{Godby_CPC07}. 
When $n\neq n'$ and $m\neq m'$ we can use the expression
Eq. (\ref{Eq:density_matrix}) to cancel the divergence; while for $n = n'$ or $m = m'$, the singularity in the Coulomb kernel is integrated out analytically, following Ref. \onlinecite{Louie_B86_GW}, around a sphere centered at $q=0$. We have also adopted another scheme using an auxiliary periodic function with the same singularity as the exact function but which can be evaluated analytically\cite{Gigi_B86}. 
These two schemes give essentially the same results. 

The eigenstates and eigenvalues of the BSE Hamiltonian provide a spectral representation of the four-point density response function (see Appendix \ref{app.bse}),
\begin{equation}
\chi_{SS'}^{\text{4P}}(\ve q,\omega)=\sum_{\lambda \lambda^{\prime}}
\frac{A_{S}^{\lambda}(\ve q) [A_{S'}^{\lambda'}(\ve q)]^{\ast} N^{-1}_{\lambda \lambda^{\prime}} }{\w-E^{\lambda}(\ve q) +i\eta}
\end{equation}
where $N_{\lambda \lambda^{\prime}}$ is the overlap matrix defined as
\begin{equation}
 N_{\lambda \lambda^{\prime}} \equiv \sum_{S}
[A_{S}^{\lambda}(\ve q)]^{\ast} A_{S}^{\lambda'}(\ve q).
\end{equation}
Using the plane wave representation (\ref{Eq:pair_density}) of the electron-hole basis states we obtain the following expression for the response function in reciprocal space  
\begin{align}
\label{Eq:chi_eh_to_GG}
\chi_{\ve G \ve G^{\prime}}(\ve q, \w) 
 = \frac{1}{\Omega} \sum_{SS^{\prime}}
\chi^{\text{4P}}_{SS^{\prime}} (\ve q, \w)  n_{S}(\ve G)
n^{\ast}_{S^{\prime}}(\ve G^{\prime}) 
\end{align}
From this expression the inverse dielectric constant and macroscopic dielectric constant follows from Eq. (\ref{Eq:epsilon_-1}) and (\ref{Eq:epsilon_M}), respectively. 

We note that upon excluding the $1/q^2$ term in the e-h exchange term,
i.e. replacing $V$ by $\bar V$ in the kernel (\ref{eq.kernel}), the
eigenstates and eigenvalues of the BSE Hamiltonian provides a spectral
representation of the irreducible response
function\footnote{Diagramatically, the irreducible response function
  is here defined as the sum of all diagrams that cannot be split into
  two by cutting an interaction line carrying momentum $\ve q$
  (diagrams which can be split into two by cutting an interaction line
  of momentum $\ve q +\ve G$) contributes to the irreducible response
  function.} rather than the full response function. In this case the
effect of $\bar V$ is to account for local field effects. Consequently
the macroscopic dielectric function can be written
\begin{align}
\label{Eq:epsilon_chi_bar}
& \epsilon(\w) = 1 - \frac{4\pi}{|\ve q|^2} \bar{\chi}_{00}(\ve q\rightarrow 0, \w) \\
& = 
1 -  \frac{4\pi}{\Omega|\ve q|^2} 
\sum_{SS^{\prime}}
 n_{S}(0) n^{\ast}_{S^{\prime}}(0) f_{S'} 
\sum_{\lambda \lambda^{\prime}}
\frac{\bar A_{\lambda}^{S}(\ve q) [\bar A_{\lambda^{\prime}}^{S^{\prime}}(\ve q)]^{\ast} \bar N^{-1}_{\lambda \lambda^{\prime}}(\ve q) }{\w-E_{\lambda}(\ve q) +i\eta}  \nonumber
\end{align}
In the above expression the optical limit $\ve q\to 0$ is taken in the
following way. First, the BSE Hamiltonian is constructed using an e-h
basis of vertical excitations ($\ve q=0$) \emph{but} using a finite
small $\ve q$ for the Coulomb interaction $1/|\ve q+ \ve G|$ in $V$
(or $\bar V$). The same finite $\ve q$ is then used when evaluating
the dielectric function from the spectral representation of the
(irreducible) response function.

\subsection{Quasiparticle energies from GLLBSC}

The derivative discontinuity $\Delta_{\mathrm{xc}}$ is defined as the difference between the fundamental gap $E_{\mathrm{g}}$ and the Kohn-Sham (KS) single-particle gap $E_{\mathrm{g}}^{\mathrm{KS}}$ as follows
\begin{equation}
 E_{\mathrm{g}} = I - A = E[n_{N-1}] - 2E[n_N]
+E[n_{N+1}] =  E_{\mathrm{g}}^{\mathrm{KS}} + \Delta_{\mathrm{xc}},
\end{equation}
where $E[n_N]$ is the total energy of the $N$-electron system and the fundamental band gap $E_\mathrm{g}$ is defined as the difference betweeen the ionization energy $I$ and the electron affinity $A$.   

Within the GLLBSC method, the derivative discontinuity $\Delta_{\mathrm{xc}}$ is obtained through 
\begin{equation}
 \Delta_{\mathrm{xc}} = \langle \Psi_{N+1} | \Delta (\ve r) | \Psi_{N+1} \rangle
\end{equation}
where
\begin{equation}\label{eq.delta}
 \Delta (\ve r) = \sum_{i}^{\mathrm{occ}} K_x \left[\sqrt{\epsilon_{\mathrm{LUMO}} - \epsilon_i}
- \sqrt{\epsilon_{\mathrm{HOMO}} - \epsilon_i} \right] 
\frac{|\psi_i(\ve r)|^2}{n(\ve r)}.
\end{equation}
 $\epsilon_i$, $\psi_i(\ve r)$ and $n(\ve r)$ are eigenvalues, eigenstates and electron density, respectively, obtained from solving the KS equation with the following GLLBSC potential
\begin{align}\label{Eq:gllb_potential}
 v_{\mathrm{GLLBSC}}(\ve r) & = 2\epsilon_{\mathrm{xc}}^{\mathrm{PBEsol}}(\ve r)  \\ 
 & + \sum_{i}^{\mathrm{occ}} K_x \sqrt{\epsilon_r - \epsilon_i} 
\frac{|\psi_i(\ve r)|^2}{n(\ve r)} + v_{\mathrm{c, resp}}^{\mathrm{PBEsol}}(\ve r) \nonumber
\end{align}
Here, $K_x \approx 0.382$ is a coefficient fitted from electron gas calculations to reproduce the exchange potential for uniform electron density and $\epsilon_r$ is a reference energy taken from the highest occupied eigenvalue. 
The GLLBSC method is an orbital dependent simplification of the KLI approximation to the exact-exchange optimized effective-potential method following the guidelines of GLLB\cite{gllb} for the exchange potential. For the details of the formulation we refer the reader to Ref. \onlinecite{gllbsc}.

\section{Implementation}
The TDDFT and BSE codes are implemented in $\textsc{gpaw}$\cite{GPAW, GPAW_10, Jun_response}, a real-space electronic structure code using the projector augmented wave methodology\cite{Blochl_B94, Blochl_03}. In this section, we focus on the construction of the screened Coulomb interaction kernel $W$, which is the most challenging and time consuming part in the BSE formalism. 
For the details of the implementation on the GLLBSC potential and the linear density response function in the PAW formalism,  we refer to Ref. \onlinecite{gllbsc} and  \onlinecite{Jun_response}, respectively.  

\subsection{Screened Coulomb interaction W}
The electron-hole correlation kernel Eq.
(\ref{Eq:W_SS}) contains the dynamically screened Coulomb interaction in a plane wave representation,
\begin{equation}
 W_{\ve G \ve G^{\prime}} (\ve q, \w) = \frac{4\pi \epsilon^{-1}_{\ve G \ve G^{\prime}}(\ve q, \w) }{|\ve q + \ve G||\ve q + \ve G^{\prime}|}. 
\end{equation}
In Eq. (\ref{Eq:W_SS}) the $\ve q$ vector represents the difference between
two $\ve k$-points in the first Brillouin zone. Thus, the $\ve q$-point mesh
has the same form as the $\ve k$-point mesh. In addition, the $\ve
q$-point mesh always includes the $\Gamma$ point, while the $\ve
k$-point mesh does not necessarily. The use of $\ve k$-point symmetry for obtaining
the wave-functions at $\ve k$-points outside the irreducible Brillouin zone has 
been described in a previous paper\cite{Jun_response}.  In the
following we describe how symmetry considerations can be used to reduce the $\ve q$-point sum.

We start by examining the $\ve q$-point symmetry in the charge density matrix defined in Eq. (\ref{Eq:pair_density}). Consider a $\ve q$ satisfying
\begin{equation}
 \ve q = T \ve q_{\mathrm{IBZ}} + \ve G_0
\end{equation}
where $\ve q_{\mathrm{IBZ}}$ is an irreducible $\ve q$ point, $T$ is a crystal symmetry transformation, and $\ve G_0$ is a reciprocal lattice vector that translates the $T \ve q_{\mathrm{IBZ}} $ vector back into the Brillouin zone if needed. The charge density matrix in Eq. (\ref{Eq:pair_density}) then becomes
\begin{align}
 & n_{n\ve k,\np \kq}(\ve G) \nonumber \\ 
&=
\langle \psi_{n\kk} | e^{-i ( T \ve q_{\mathrm{IBZ}} + \ve G_0 + \ve G) \cdot \rr} | \psi_{n^{\prime} \ve k +\ve q}
\rangle  \nonumber \\
&= 
\langle \psi_{nT^{-1}\kk} | e^{-i [\ve q_{\mathrm{IBZ}} + T^{-1}(\ve G_0 + \ve G)] \cdot \rr} | \psi_{n^{\prime} T^{-1}(\ve k +\ve q)}
\rangle \nonumber \\ 
&= n_{n T^{-1}\ve k, n^{\prime} T^{-1}(\ve k+\ve q)}(T^{-1}(\ve G_0 + \ve G))
\end{align}

Since the calculation of $\chi^0_{\ve G \ve G^{\prime}} (\ve q, \w)$ involves the summation of the charge density matrix over all the BZ $\ve k$-points, the above equation leads directly to the following relation (as long as  $T^{-1}\ve k$ belongs to the $\ve k$-point mesh): 
\begin{equation}
  \chi^0_{\ve G \ve G^{\prime}} (\ve q, \w) = 
\chi^0_{T^{-1}(\ve G + \ve G_0), T^{-1}(\ve G^{\prime}+ \ve G_0)} (\ve q_{\mathrm{IBZ}}, \w). 
\end{equation}
The above relation also applies to $W_{\ve G \ve G^{\prime}} (\ve q, \w)$.

Besides crystal symmetry, time reversal symmetry is also used for systems that have no inversion symmetry. If the transformation of a given $\ve q$ to IBZ requires both crystal symmetry and time reversal symmetry via  
\begin{equation}
 \ve q = - T \ve q_{\mathrm{IBZ}} + \ve G_0,
\end{equation}
the $W$ matrix should satisfy
\begin{equation}
   W_{\ve G \ve G^{\prime}} (\ve q, \w) = 
W^{\ast}_{-T^{-1}(\ve G + \ve G_0), -T^{-1}(\ve G^{\prime}+ \ve G_0)} (\ve q_{\mathrm{IBZ}}, \w). 
\end{equation}

Finally, it has to be emphasized that for a finte $\ve k$-point mesh used in a numerical calculation, the crystal symmetry transformation $T$ should apply to both $\ve q$-points and $\ve k$-points. This results in reduced crystal symmetry operations if the $\Gamma$ centered $\ve q$-point mesh does not coincide with the $\ve k$-point mesh.

\section{Solids}
In this section the optical properties of a representative set of six
bulk semiconductors and insulators are studied using both ALDA and
the BSE. We start by presenting the fundamental gaps obtained with LDA
and GLLBSC. The accuracy of the GLLBSC gaps is similar to
G$_{0}$W$_{0}$ calculations from the litterature with an average
absolute deviation of 0.3 eV from experiments. An important ingredient
in the BSE calculation of optical spectra is the static dielectric
constant which determines the strength of the screened electron-hole
interaction, $W$. We find that the best agreement with experiment is
obtained when the response function is evaluated from the LDA or
GLLBSC Kohn-Sham (i.e. without adding the derivative
discontinuity)energies, and we explain this from the fact that the
electron-hole interaction is not explicitly accounted for by the
random phase approximation used to obtain $\epsilon$. Finally, the
absorption spectra using both ALDA and BSE are presented. Very good
agreement with the experimental spectra is found for the GLLBSC-BSE
combination both for the absorption onset and the excitonic features.

\begin{table}[t]
\caption{Band gaps (units in eV) calculated using GLLBSC without (wo.) and with (w.) the derivative discontinuity $\Delta_{\mathrm{xc}}$ added to the Kohn-Sham gap. These values are compared with LDA, $G_0 W_0$ and experimental data. Underlined values correspond to zero-temperature values. The mean absolute errors (MAE) with respect to experiments are summarized in the last row.}

\begin{tabular*}{0.95\linewidth}{@{\extracolsep{\fill}} l c c c c c}
  \hline \hline
  & LDA & GLLBSC & GLLBSC & $G_0 W_0$ & Expt.  \\
  &     &  (wo.) & (w.)   &           \\
\hline
Si      &  0.51  &  0.74 & 1.09   & 1.12\footnotemark[1]  & \underline{1.17}\footnotemark[2] \\
C       &  4.16  &  4.22 & 5.52  & 5.50\footnotemark[1]   & 5.48\footnotemark[3] \\
InP     &   0.61  &  1.15 & 1.63  & 1.32\footnotemark[4]   & \underline{1.42}\footnotemark[2]   \\
MgO     &  4.63  &  6.10 & 8.32 & 7.25\footnotemark[1]   & 7.83\footnotemark[5] \\
GaAs  &   0.57    & 0.93  & 1.23 & 1.30\footnotemark[1]   & \underline{1.52}\footnotemark[2] \\
LiF     &  8.87  & 10.97 & 14.94 & 13.27\footnotemark[1]  & 14.20\footnotemark[6] \\
\hline
MAE & 2.04 & 1.25 & 0.31 & 0.32 \\
  \hline \hline 
\end{tabular*}
\footnotetext[1]{Reference \onlinecite{G0W0_bandgap}}
\footnotetext[2]{Reference \onlinecite{Kittel}, T=0K}
\footnotetext[3]{Reference \onlinecite{Cardona}}
\footnotetext[4]{Reference \onlinecite{InP_GW_bandgap}}
\footnotetext[5]{Reference \onlinecite{MgO_bandgap}}
\footnotetext[6]{Reference \onlinecite{LiF_bandgap}}
\end{table}

\subsection{Fundamental gaps}

Table I shows the calculated band gaps for Si, C, InP, MgO, GaAs and
LiF. We have used the experimental lattice
constants for all systems: Si (5.431 \AA), C (3.567 \AA), InP (5.869 \AA), MgO
(4.212 \AA), GaAs (5.650 \AA) and LiF (4.024 \AA). The Kohn-Sham energies and wave functions were obtained with GPAW using uniform grids with
spacing 0.2 \AA~ and a Fermi temperature of 0.001 eV. The Brillouin zone was sampled using a  
Monkhorst-Pack grid of $24 \times 24 \times 24$ which was found sufficient to converge the band gaps to within 0.02 eV.

\begin{table}[t]
\caption{The static macroscopic dielectric constant $\epsilon$ obtained using TDDFT on top of LDA as well as GLLBSC electronic structure without (wo.) and with (w.) discontinuity $\Delta_{\mathrm{xc}}$ applied.  The two rows for each semiconductor correspond to TDDFT calculations with RPA and the ALDA kernel, respectively.   }
\begin{tabular*}{0.95\linewidth}{@{\extracolsep{\fill}} l c c c c }
  \hline \hline
 & \multicolumn{1}{c}{LDA}  & \multicolumn{1}{c}{GLLBSC}  &  \multicolumn{1}{c}{GLLBSC}  & Expt.  \\
 & \multicolumn{1}{c}{}  & \multicolumn{1}{c}{(wo.)}  &  \multicolumn{1}{c}{(w.)}  &   \\
 \hline
 Si (RPA)     & 12.53   & 11.00    &   10.25    &   11.90\footnotemark[7]    \\
  \ \  (ALDA)   & 13.16  & 11.54 &  10.73   &       \\ 
  C             &  5.56   &  5.48      &   5.04    &  5.70\footnotemark[7]  \\
                &   5.82   &  5.74     &   5.25     &  \\
 InP           &   11.48  &  8.92     &  8.06 & 12.5\footnotemark[7] \\
                &   11.99  &  9.33     &  8.41 & \\
 MgO         &   3.06   &  2.52    & 2.31  &  2.95\footnotemark[8]\\
                &   3.20   &  2.63    &  2.39  &  \\
 GaAs        &   13.52  &  11.12  &  10.28   &  11.10\footnotemark[7] \\
               &    14.17  &   11.68  & 10.78   & \\
  \hline \hline 
\end{tabular*}
\footnotetext[7]{Reference \onlinecite{dielectric_constant_handbook}, T=300K.}
\footnotetext[8]{Reference \onlinecite{MgO_optical_epsilon}, optical dielectric constant.}
\end{table}

\begin{figure*}[t]
    \centering
    \includegraphics[width=1.\linewidth,angle=0]{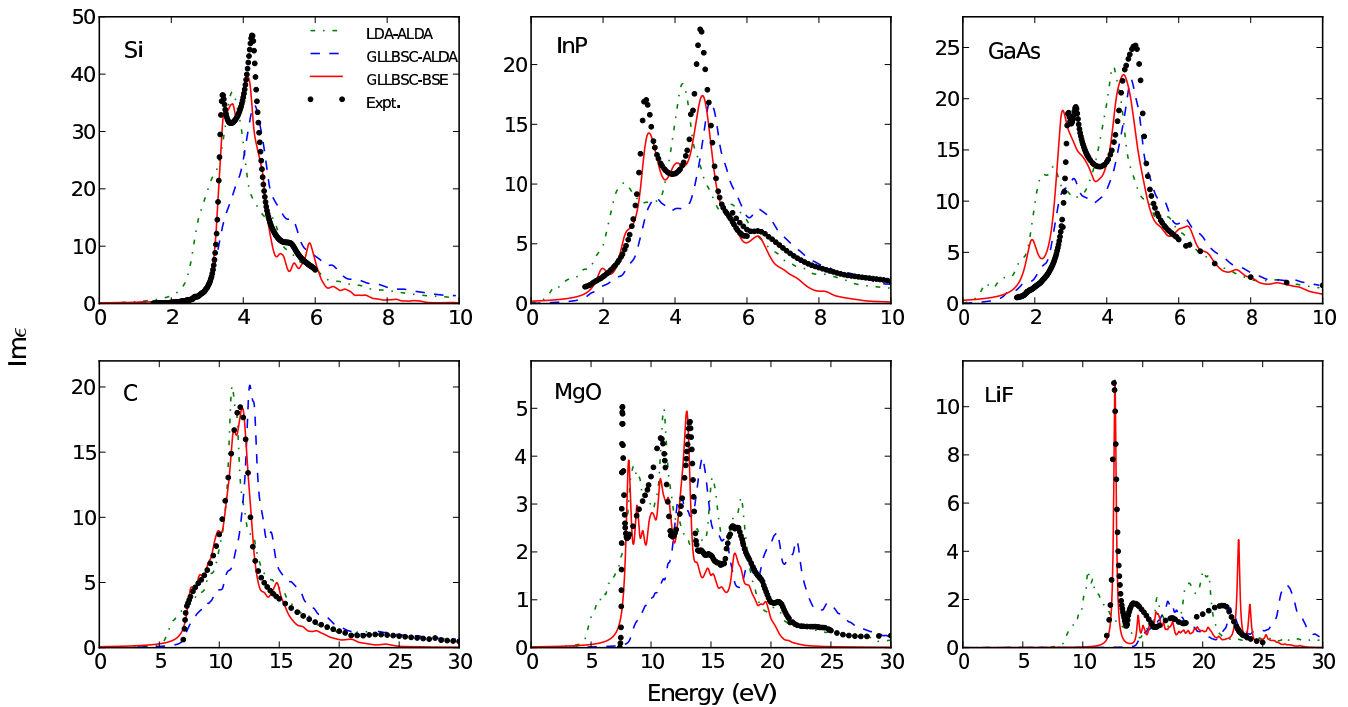}
    \caption{\label{Fig:absorption}Optical absorption spectra calculated using LDA-ALDA (dash-dotted line), GLLBSC-ALDA (dashed line) as well as GLLBSC-BSE (solid line).  The derivative discontinuity, $\Delta_{\mathrm{xc}}$, is included in the GLLBSC calculations. The calculated spectra are compared with experimental data (dots, Ref. \onlinecite{dielectric_function_handbook}). }
\end{figure*}

Compared to LDA band gaps (first column), GLLBSC even without
the discontinuity (second column) improves the band gaps. The reason is
that the GLLBSC potential Eq. (\ref{Eq:gllb_potential}) can
reproduce the asymptotic $1/r$ behavior of the Coulomb
potential\cite{gllb} and thus the Kohn-Sham eigenvalues are improved over LDA. 
By adding the discontinuity (third column), the
band gaps agree reasonably well with experimental data (last column).
The mean absolute error (MAE) with respect to the experimental data is 0.31 eV in agreement with a previous study using GLLBSC for oxides in the perovskite structure\cite{Ivano}.  The sign of the deviations from experiment seem to vary randomly. This is in contrast to the G$_0$W$_0$ results (fourth
column)\footnote{These $G_0 W_0$ results were obtained using plane wave basis and the PAW method and are thus directly comparable to our results.}, which systematically
underestimates the band gaps with the largest error being almost 1 eV.
We note that (quasi-) selfconsistent GW calculations have been shown to improve the ionization potentials of molecules\cite{Rostgaard_B10} and band gaps of solids\cite{Schilfgaarde_L06} by reducing the overscreening resulting from the LDA starting point. However, such calculations are even more computationally demanding than G$_0$W$_0$, and are therefore not normally used for the calculation of optical spectra. We will show in the
following that GLLBSC represents a cheap alternative means to GW providing not only
reasonable fundamental gaps, but also very good optical dielectric constants
and absorption spectra.

\subsection{Dielectric constants}

Table II shows the calculated static macroscopic dielectric constants.
In addition to the parameters presented for obtaining the band gaps,
60 - 90 unoccupied bands, corresponding to around 140 eV above
the Fermi level, were used in the calculation of the response
function Eq. (\ref{Eq:chi0_GG}). Local field effects were included up to an energy cutoff of
150 - 250 eV, which varies according to the size of the unit cell and
corresponds to 169 $\ve G$ vectors.  The static dielectric constants
obtained using LDA-RPA (first column), that is, RPA calculations based
on LDA wave functions and energies, are generally higher than the experimental
values (last column) due to the underestimated LDA band gaps.  The
overestimation is enhanced by inclusion of the ALDA kernel (the second row
for each semiconductor), in agreement with previous
studies\cite{Jun_response}. The GLLBSC without the discontinuity increase the band gaps relative to LDA and consequently reduces the dielectric function towards the experimental value. The inclusion of the
discontinuity further opens up the gap and the corresponding dielectric constants (third column) systematically underestimate the experimental values.
 This underestimation is a result of the neglect of electron-hole interaction when the response function is evaluated at the RPA and (to some extent) ALDA levels. In order to reduce the error coming from this effect, the response function should be evaluated using "dressed" single-particle energies rather than the bare QP energies.
In the following, we use the GLLBSC(wo.)-RPA dielectric function for calculating $W$.

\subsection{Absorption spectra}
The absorption spectra calculated using TDDFT and the BSE are shown
in Fig. 1. TDDFT calculations were performed using the ALDA kernel
and the same parameters as used for obtaining the dielectric constants (see previous section). For the BSE calculations we used an $8
\times 8 \times 8$ Monkhorst-Pack $k$-point grid not containing the Gamma-point (for InP $10
\times 10 \times 10$ $k$-points were used).  We have also checked the
spectra with $12 \times 12 \times 12$ $k$-point sampling. The main
peaks in the absorption spectra are well converged with the applied $k$-point sampling, however, a complete elimination of the small "wiggles" seen in the spectra
would require significantly denser $k$-point sampling. The screened interaction
kernel, $W_{\ve G \ve G^{\prime}}(\ve q)$, was obtained using GLLBSC(wo.)-RPA,
with 60 unoccupied bands and local field effects included by 169 $\ve G$-vectors. Three valence and three conduction
bands were taken into account in constructing the BSE matrix. Again, this is sufficient to converge the major (excitonic) peaks and the low energy part of the absorption spectra.
The Tamm-Dancoff approximation\cite{Rubio_RMP02}, consisting of the neglect of coupling between v-c and c-v transitions, was employed. The effect of 
temperature, which in general lowers the band gap and smears the
absorption spectrum\cite{Marini_L08}, is not considered in the current
work. As a result, the spectra presented here are broadened using
smearing factors (in units of eV): Si (0.10), C (0.35), InP (0.20),
MgO (0.25), GaAs (0.20) and LiF (0.12). 

\begin{figure}[t]
    \centering
    \includegraphics[width=1.\linewidth,angle=0]{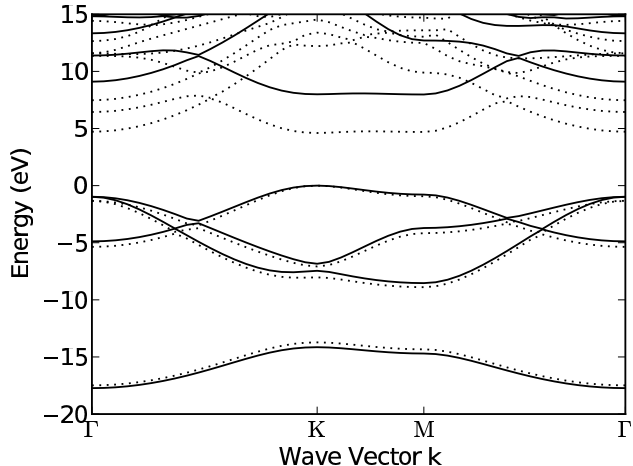}
    \caption{\label{Fig:BN_bandstr}Band structure of a h-BN sheet calculated with GLLBSC (solid lines) and LDA (dotted lines). The top of the valence bands is set to zero. }
\end{figure}

As can be seen from the absorption spectra in Fig. \ref{Fig:absorption}, LDA-ALDA (green dash-dotted lines) gives threshold optical transition energies that are 0.5 - 3 eV lower than experiments (black
dots). This is a result of the too low LDA band gaps. The use of GLLBSC wave functions and energies including the derivative discontinuity, GLLBSC-ALDA (blue dashed lines) increases the absorption threshold energies and improves the agreement with experiments. However, the
shape of the spectra are qualitalitvely different. In particular, the spectra are too low at the on-set of the absorption and the excitonic features in Si, MgO, and LiF, are completely missed. This is because ALDA does not properly account for electron-hole
interactions. In contrast the spectra obtained from the BSE using the GLLBSC eigenvalues as QP energies (red lines) are in excellent agreement with experiments.
A small exception is for GaAs where a small peak, abscent in the experimental spectrum, is seen at around 2 eV. A similar feature was seen 
in a previous calculaton employing a non-local approximation to the xc kernel within
TDDFT\cite{Gross_L11}, but does not appear in a previous GW-BSE
calculation\cite{Alouani_B01}. This indicates that the presence of the feature is related to differences between the GLLBSC and GW band structure.
We note that (small) deviations between the GLLBSC and GW band structures was recently proposed as the reason for (slight) inaccuracies in the GLLBSC-ALDA calculated surface plasmon energies of Ag(111)\cite{Jun_AgH}.

\section{Graphene/boron-nitride}
In this section we study the bandstructure and optical absorption
spectra of graphene, a single layer of hexagonal boron-nitride (h-BN),
and their interface graphene/h-BN. The lattice parameter of h-BN is very similar to that of graphene making it a
promising candidate substrate material for graphene based
devices\cite{NN_2010}.  In contrast to graphene, which is a
semi-metal, h-BN has a wide band gap and exhibits strong excitonic
effects. The optical properties of layered BN sheets as well as BN
nanotubes have been studied extensively both
experimentally\cite{NM_2004} and
theoretically\cite{Alouani_L06_bulkBN, Rubio_L96}. Upon adsorption of
graphene onto a h-BN sheet, a small bandgap of around 10-200 meV, depending on
the configuration and interplane distance, emerges\cite{Kelly_B07}. The ground state electronic properties, including the role of dispersive forces, and the band structure have been studied\cite{Kelly_B07, Lichtenstein_B11}. Below we investigate the optical properties of the graphene/h-BN interface and assess the quality of the GLLBSC for such 2D structure.

Before presenting the results for graphene/h-BN, first we examine a single h-BN
sheet. For the lattice constant of h-BN we used 2.89\AA~and 20\AA~vacuum was included between the periodically repeated BN layers. 
Figure \ref{Fig:BN_bandstr} shows the band structure calculated using LDA (dotted lines) and GLLBSC (solid lines).  The LDA band gap (situated at the K-point) is 4.61 eV which is 0.3 eV larger than reported in an earlier pseudopotential study\cite{Rubio_B95}. The GLLBSC band gap is 7.99 eV, which
includes the derivative discontinuity of 2.12 eV, is close to the pseudopotential G$_0$W$_0$ band gap of 7.9 eV\cite{Rubio_L96}.

\begin{figure}[t]
    \centering
    \includegraphics[width=1.\linewidth,angle=0]{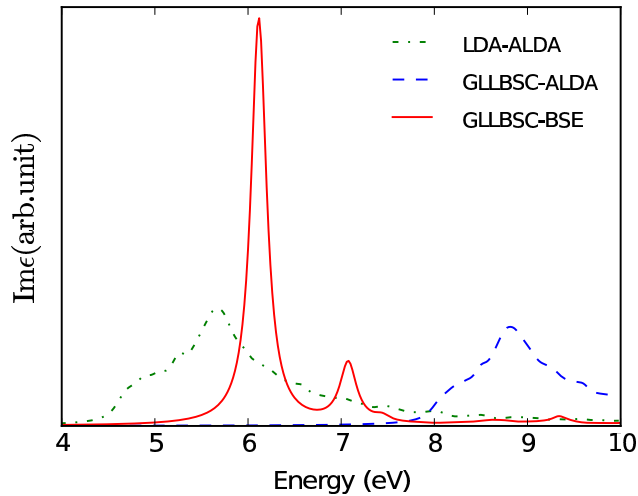}
    \caption{\label{Fig:BN_bse}Optical absorption spectra of a h-BN sheet calculated using LDA-ALDA (dash-dotted line), GLLBSC-ALDA (dashed line) and GLLBSC-BSE (solid line). }
\end{figure}

Figure \ref{Fig:BN_bse} shows the absorption spectrum of a h-BN sheet
obtained with three different methods. The LDA-ALDA spectrum shows a
broad absorption peak with an onset at 4.5 eV in good agreement with
literature\cite{Rubio_L96}. The GLLBSC-ALDA spectrum is essentially
identical to LDA-ALDA, but blue shifted by the difference in the band
gap.  For the BSE calculation, the Brillouin zone was sampled on a non
Gamma-centered $32\times 32$ Monkhorst-Pack grid, and 70 unoccupied
bands were included to obtain the screened interaction $W$. A
two-dimensional Coulomb cutoff technique\cite{Rubio_cutoff} was used
to avoid interactions between supercells. Since we are interested in
the low-energy part of the absorption spectrum and because the valence
and conduction bands are well separated from the rest of the bands in
the relevant part of the Brillouin zone (around the K-point), only the
valence and conduction bands were included in the BSE effective
Hamiltonian. The absorption spectrum obtained with GLLBSC-BSE shows
three excitonic peaks at 6.1, 7.1 and 7.4 eV with decreasing
amplitude. These exciton energies agree well with the value of 6.2 eV,
7.0 eV and 7.4 eV obtained with the GW-BSE scheme\cite{Rubio_L96}.

\begin{figure}[t]
    \centering
    \includegraphics[width=1.\linewidth,angle=0]{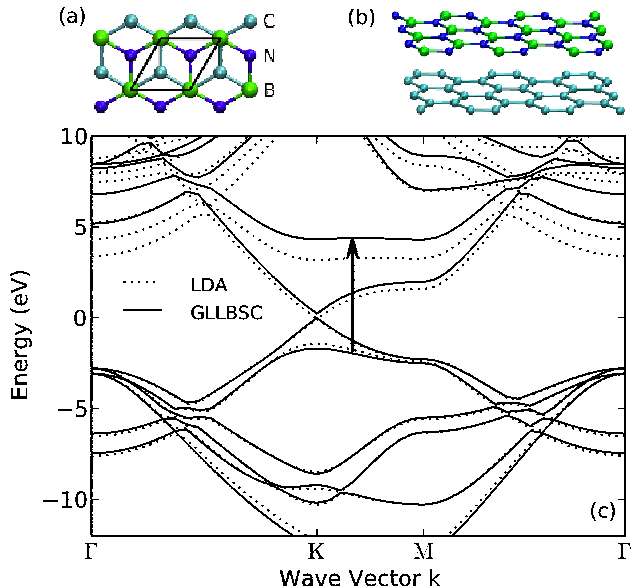}
    \caption{\label{Fig:G_BN_bandstr} Top (a) and side (b) view of a graphene/h-BN. (c) Band structure of graphene/h-BN calculated with GLLBSC (solid lines) and LDA (dotted lines). The top of the valence bands is set to zero.}
\end{figure}

For the graphene/h-BN interface, we studied the structure where one C
atom is ontop of a B atom and the other C atom is above the center of
the BN ring, as shown in Fig. \ref{Fig:G_BN_bandstr} (a) and (b). Both graphene and h-BN are kept planar at a distance 3.48\AA~apart. Recent RPA calculations found this structure and adsorption
distance to be the most stable\cite{Lichtenstein_B11}. Fig.
\ref{Fig:G_BN_bandstr} shows the band structure of graphene/h-BN. For
the LDA band structure (dotted lines), a small band gap of 31 meV
opens at the K point. This number is very close to the 53 meV found in
an earlier study\cite{Kelly_B07}. The h-BN gap, indicated by the arrow
and is 4.60 eV in the LDA, which is essentially the same as found for
the isolated h-BN sheet (4.61 eV). This is in contrast to the GLLBSC
band structure which yields a band gap of the adsorbed h-BN of 6.01 eV
which is 1.98 eV lower than obtained for isolated h-BN. This sizable
reduction of the gap is not due to hybridization, but rather is a
result of a reduction of the derivative discontinuity from 2.12 eV to
essentially zero. We note in passing that the GLLBSC value of 6.01 eV is 0.3 eV larger than our G$_0$W$_0$ results for this system (to be published elsewhere).

\begin{figure}[t]
    \centering
    \includegraphics[width=1.\linewidth,angle=0]{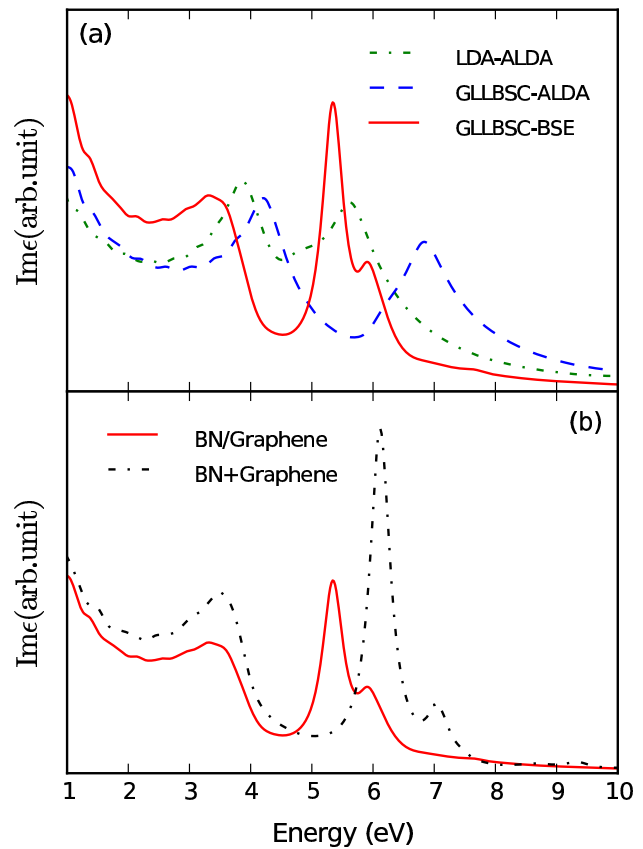}
    \caption{\label{Fig:G_BN_bse}Upper panel: Optical absorption spectrum of graphene/h-BN calculated using LDA-ALDA (dash-dotted line), GLLBSC-ALDA (dashed line) and GLLBSC-BSE (solid line). Lower panel: The GLLBSC-BSE spectrum of the interface (repeated) together with the sum of the absorption spectra of an isolated graphene and BN layer, respectively.}
\end{figure}

The reduction of the fundamental gap when BN is adsorbed on graphene
is physically meaningful and can be explained by the screening
provided by the graphene layer (image charge effect) which reduces the energy cost of removing electrons/holes from the BN layer. For molecules on
surfaces, this effect has been shown to be well described by the GW
method, whereas both (semi-)local and hybrid functionals completely
miss the effect predicting no change in the gap upon adsorption (apart from obvious hybridization effects)\cite{Neaton_L06, Juanma_B09}. Interestingly, within the GLLBSC the gap reduction is a result of the vanishing, or strong reduction, of the derivative
discontinuity. However, this also has the unphysical consequence that the
reduction is present independent of the graphene-BN distance. This follows from the obsrvation that the derivative discontinuity in Eq. (\ref{eq.delta}) becomes
zero for a metallic system.

The absorption spectrum of graphene/BN calculated with the three
different schemes are shown in Fig. \ref{Fig:G_BN_bse}(a). Due to the
semi-metallic nature of graphene and the dense set of intra band
transitions in the 0-5 eV energy region, a much denser k-point
sampling is required to obtain a smooth absorption spectrum for this
system. We used a $80\times 80$ Monkhorst-Pack grid for both
the ALDA and BSE calculations.  70 unoccupied bands were taken into
account for the calculation of the response function, while 2 valence
and 2 conduction band were included in the BSE Hamiltonian. The energy
range below 1 eV is not shown in the figure since the excitations
close to the Dirac point requires even denser k-points sampling.

The LDA-ALDA spectrum (dashed-dotted line) shows absorption peaks at
3.9 and 5.6 eV originating from transitions within the graphene and BN
layer, respectively.  It closely resembles a superposition
of the spectra from freestanding graphene (not shown here) and BN
sheets (dashed-dotted line in Fig. \ref{Fig:BN_bse}) , with only a
minor difference of 0.1 eV in peak positions. Using GLLBSC-ALDA
(dashed line), the two peaks shift up to 4.2 and 6.8 eV, respectively.
The shift in the BN peak position is in accordance with the shift in
the BN gap in Fig. \ref{Fig:G_BN_bandstr}. Note that the graphene peak
energy of 4.2 eV is much lower than the 5.15 eV obtained from a
previous G$_0$W$_0$ calculation (without electron-hole
interaction)\cite{Louie_L09_graphene}. We speculate that the deviation
is due to an incorrect description of the slope of the graphene bands
around the Dirac point where GLLBSC yields essentially the LDA result,
see Fig. \ref{Fig:G_BN_bandstr}.  Although the absolute absorption peak
for graphene is underestimated, the excitonic effect is still well
described using the BSE. With electron-hole pair interaction included
(solid line), the graphene absorption peak at 4.2 eV is redshifted by
0.6 eV, the same amount as was found in Ref.
\onlinecite{Louie_L09_graphene}. The shift in the BN peak is, however,
more striking. Upon adsorption of graphene, the BN exciton peak shifts
from 6.1 eV in Fig. \ref{Fig:BN_bse} to 5.4 eV in Fig.
\ref{Fig:G_BN_bse}. The reduction of the exciton energy of 0.7 eV is
much smaller than the 1.98 eV reduction of the fundamental gap. This
means that the exciton binding energy has been reduced from 1.9 eV in
freestanding BN to 0.6 eV when adsorbed on graphene. Again, this is
explained by the enhanced screening of the electron-hole pair provided
by the electrons in graphene. The substrate induced screening of
exciton binding energies was recently observed in GW-BSE calculations
for molecules adsorbed on a metal surface.\cite{Juanma_L11}

\section{Conclusions}
We have presented an implementation of the Bethe-Salpeter equation (BSE) which allows for the calculation of optical properties of
materials with proper account of electron-hole interactions. Rather
than following the standard approach where quasiparticle energies are
obtained from the computationally costly GW method, we showed that
excellent agreement with experimental absorption spectra of a representative set 
of semiconductors and insulators, can be obtained by using
single-particle energies from the GLLBSC functional. The latter yields
very good fundamental gaps due to its explicit inclusion of the
derivative discontinuity, and its computational cost is comparable to
LDA. For  a single layer of boron-nitride the fundamental gap and
optical spectrum obtained with GLLBSC-BSE is very close to that of
previous GW-BSE calculations. We showed that when BN is adsorbed on
graphene, the fundamental gap is reduced by 2 eV. This reduction can
be explained by image charge screening, and shows up in the GLLBSC
calculation as a vanishing contribution from the derivative
discontinuity. Finally, we found that the absoption spectrum of
graphene/BN interface is not simply a sum of the absorption spectra of
the isolated layers, because the transition energies in BN become
redshifted by up to almost 1 eV due to screening by the graphene electrons.

\appendix
\section{Effective two-particle Hamiltonian}\label{app.bse}
To obtain an effective two-particle Hamiltonian describing the optical excitations of the interacting electron system, we begin by considering the Bethe-Salpeter equation (BSE) for the (retarded) four-point response function, $\chi^{\text{4P}}$.
Assuming a static electron-hole interaction kernel $\chi^{\text{4P}}$ can be written
\begin{widetext}
\begin{equation}
\label{Eq:Dyson_4point}
 \chi^{\text{4P}}(\ve r_1 \ve r_2 ;\ve r_3 \ve r_4,\omega)  = P^{\text{4P}}(\ve r_1 \ve r_2; \ve r_3 \ve r_4,\omega) + \int P^{\text{4P}}(\ve r_1 \ve r_2; \ve r_5 \ve r_6,\omega) 
K^{\text{4P}}(\ve r_5 \ve r_6 ;\ve r_7 \ve r_8)\chi^{\text{4P}}(\ve r_7 \ve r_8 ;\ve r_3 \ve r_4,\omega) d \ve r_5 d \ve r_6 d \ve r_7 d \ve r_8  
\end{equation}
\end{widetext}
In writing the above BSE equation we have made the simplifying, and for
practical purposes essential, assumption that the electron-hole
interaction kernel, $K$, is frequency independent. The quantity
$\chi^{\text{4P}}$ is an uncontracted version of the density response
function, i.e. $\chi(\ve r,\ve r',\omega)=\chi^{\text{4P}}(\ve r \ve r
;\ve r' \ve r',\omega)$ while $P^{\text{4P}}$ is the four-point
response function for independent (but self-energy dressed)
quasiparticles (QP). The kernel is given by $K^{\text{4P}}=V-\frac{1}{2}W$ where
\begin{equation}
V(\ve r_1 \ve r_2 ;\ve r_3 \ve r_4) =\frac{1}{|\ve r_1 - \ve r_3|} \delta(\ve r_1-\ve r_2)\delta(\ve r_3- \ve r_4)
\end{equation}
is the electron-hole exchange and 
\begin{equation}
W(\ve r_1 \ve r_2 ;\ve r_3 \ve r_4) =\int \frac{\epsilon^{-1}(\ve r_1,\ve r',0)}{|\ve r' - \ve r_2|}d \ve r' \delta(\ve r_1-\ve r_3)\delta(\ve r_2 - \ve r_4)
\end{equation}
is the statically screened direct electron-hole interaction.

Assuming that the QP energies and wave functions can be described by an effective non-interacting Hamiltonian, $H_{\text{QP}}$, we can write the independent response function as   
\begin{align}\label{Eq:chi0_r}
& P^{\text{4P}}(\ve r_1 \ve r_2; \ve r_3 \ve r_4, \w) = \frac{2}{\Omega} 
 \sum_{\ve q}\sum_{\ve k n m} (f_{n\ve k}-f_{m \ve k + \ve q}) \nonumber\\
&  \times  
 \frac{\psi_{n \ve k}^{\ast}(\ve r_1) \psi_{m \ve k+\ve q}(\ve r_2)
\psi_{n \ve k}(\ve r_3) \psi_{m \ve k+\ve q}^{\ast}(\ve r_4)}{\omega +
\epsilon^{\text{QP}}_{n\ve k} - \epsilon^{\text{QP}}_{m \ve k+\ve q } + i\eta}.
\end{align} 
where the wave functions form an orthonormal set and the occupation
factors are 1 or 0 for occupied and empty states, respectively. 

The full four-point response function can also be expanded in the orthonormal basis of single-particle transitions, $\psi_S(\ve r_1,\ve r_2)=\psi_{n \ve k}^{\ast}(\ve r_1) \psi_{m \ve k+\ve q}(\ve r_2)$,
\begin{align}
\label{Eq:chi_transform}
 & \chi^{\text{4P}}(\ve r_1 \ve r_2; \ve r_3 \ve r_4, \w) 
 = \sum_{\ve q}\sum_{SS^{\prime}} 
 \chi_{SS^{\prime}}(\ve q, \w) \\
& \times \psi_{n \ve k}^{\ast}(\ve r_1) \psi_{m \ve k+\ve q}(\ve r_2)
\psi_{n' \ve k'}(\ve r_3) \psi_{m' \ve k'+\ve q}^{\ast}(\ve r_4) \nonumber
\end{align}
As a consequence of the periodicity of the crystal lattice, all 4-point functions are diagonal in $\ve q$. Note that the indices $n,m,n',m'$ must run over all bands, both occupied and unoccupied, in order to ensure that the two-particle basis is complete (it will, however, turn out that it is sufficient to consider only e-h and h-e transitions).

The non-interacting response function is diagonal in the two-particle basis, 
\begin{equation}
\label{Eq:chi0_transform}
P^{\text{4P}}_{SS^{\prime}}(\ve q,\w) = \frac{f_S}{\w-\varepsilon_S+i\eta} \delta_{S S^{\prime}}
\end{equation}
where the occupation and transition energy for an electron hole pair $S$ is defined as
\begin{align}
 f_S \equiv f_{n \ve k}-f_{m \ve k+\ve q} \\
 \varepsilon_S \equiv \varepsilon_{n \ve k}-\varepsilon_{m \ve k+\ve q}
\end{align}

The four point Bethe-Salpeter equation Eq. (\ref{Eq:Dyson_4point}) in the two-particle basis corresponding to momentum transfer $\ve q$ becomes
\begin{align}\label{Eq:Dyson_eh}
  \chi_{SS^{\prime}}^{4\text{P}}(\ve q, \w)
& = P^{4\text{P}}_{SS}(\ve q,\w) \\  
& + 
\sum_{S^{\prime\prime}}P^{4\text{P}}_{SS}(\ve q,\w)
  K^{4\text{P}}_{S S^{\prime\prime}}(\ve q,\w) 
  \chi^{4\text{P}}_{S^{\prime\prime}S^{\prime}}(\ve q,\w)
 \nonumber
\end{align}
Expressions for the kernel matrix elements are given in Eqs. (\ref{Eq:V_eh}) and (\ref{Eq:W_SS}).

Substituting Eq. (\ref{Eq:chi0_transform}) into Eq. (\ref{Eq:Dyson_eh}) and rearranging yields 
\begin{equation}\label{Eq:Dyson_eh_rearrange}
\chi_{SS^{\prime}}^{4\text{P}}(\ve q, \w)=[I (\w+i\eta)-\mathcal{H}(\ve q, \w)]^{-1}_{SS'}f_{S'}
\end{equation}
where the effective two-particle Hamiltonian, $\mathcal{H}$, is defined as
\begin{align}
  \mathcal{H}_{S S^{\prime}}(\ve q, \w) 
 \equiv 
\varepsilon_S \delta_{S S^{\prime}}  + f_S
K^{4\text{P}}_{S S^{\prime}}(\ve q, \w),
\end{align}
and $I$ is an identity matrix with the same dimension as $\mathcal{H}$.

By dividing the matrices into $4\times 4$ blocks corresponding to
two-particle basis functions containing e-h, h-e, e-e, and h-h
transitions, it follows that
$\chi_{SS^{\prime}}^{4\text{P}}$ is non-zero only within the $2\times
2$ upper left block. For this reason we can reduce the problem by limiting the two-particle basis functions, $\psi_S$, to the e-h and h-e states. Using the eigenstates and energies of the BSE Hamiltonian,
\begin{equation}\label{Eq:H_eh}
 \mathcal{H}(\ve q) A_{\lambda}(\ve q) = E_{\lambda}(\ve q) A_{\lambda}(\ve q)
\end{equation}
we can construct the spectral representation of the resolvent of the BSE Hamiltonian,
\begin{equation}\label{Eq:resolvent}
 \left[I (\w+i\eta)-\mathcal H(\ve q) \right]^{-1}_{S S^{\prime}}
=  \sum_{\lambda \lambda^{\prime}}
\frac{A_{\lambda}^{S}(\ve q) [A_{\lambda^{\prime}}^{S^{\prime}}(\ve q)]^{\ast} N^{-1}_{\lambda \lambda^{\prime}}(\ve q) }{\w-E_{\lambda}(\ve q) +i\eta}
\end{equation}
where $N_{\lambda \lambda^{\prime}}(\ve q)$ is the overlap matrix defined as
\begin{equation}
 N_{\lambda \lambda^{\prime}}(\ve q) \equiv \sum_{S}
[A_{\lambda}^{S}(\ve q)]^{\ast} A_{\lambda^{\prime}}^{S}(\ve q)
\end{equation}
The BSE Hamiltonian (\ref{Eq:H_eh}) is in general non-Hermitian as a matrix in the e-h and h-e basis. However, within the standard
Tamm-Dancoff approximation, in which only the e-h transitions are considered (i.e. transitions with positive energies), $\mathcal{H}(\ve q)$ becomes Hermitian and $N_{\lambda \lambda^{\prime}}(\ve q) = \delta_{\lambda \lambda^{\prime}}$. 

Since the two-point response function, $\chi_{\ve G \ve G'}(\ve q,\w)$, is obtained by Fourier transforming $\chi^{\text{4P}}(\ve r \ve r
;\ve r' \ve r',\omega)$, we conclude from Eq. (\ref{Eq:chi_transform}) that
\begin{align}
\chi_{\ve G \ve G^{\prime}}(\ve q, \w) 
 = \frac{1}{\Omega} \sum_{SS^{\prime}}
\chi^{4\text{P}}_{SS^{\prime}} (\ve q, \w)  n_{S}(\ve G)
n^{\ast}_{S^{\prime}}(\ve G^{\prime}) 
\end{align}
where the charge density matrix, $n_{S}(\ve G)$, is defined in Eq. (\ref{Eq:pair_density}).

Finally, the relation to the macroscopic dielectric function Eq.
(\ref{Eq:epsilon_chi_bar}) is established using Eqs.
(\ref{Eq:Dyson_eh_rearrange}) and (\ref{Eq:resolvent}), together with
the relation
\begin{equation}
\epsilon(\w) = 1 - \frac{4\pi}{|\ve q|^2} \bar{\chi}_{00}(\ve q\rightarrow 0, \w) 
\end{equation}
between the dielectric function and the irreducible response function,
$\bar \chi$. As discussed in Sec. \ref{sec.bse} the latter is obtained
in place of $\chi$ when the long range $\ve G=0$ term excluded from
the e-h exchange kernel in Eq. (\ref{Eq:V_eh}).

\begin{acknowledgments}
The Center for Atomic-scale Materials Design is sponsored
by the Lundbeck Foundation. The Catalysis for Sustainable Energy initiative is
funded by the Danish Ministry of Science, Technology and Innovation. The Center for Nanostructured Graphene is sponsored
by the Danish National Research Foundation. J. Yan acknowledges support from Center for Interface Science and Catalysis (SUNCAT) through the U.S. Department of Energy, Office of Basic Energy Sciences. 
The computational studies were supported as part of the 
Center on Nanostructuring for Efficient Energy Conversion, an Energy 
Frontier Research Center funded by the U.S.
Department of Energy, Office of Science, Office of Basic Energy 
Sciences under Award No. DE-SC0001060.

\end{acknowledgments}


\end{document}